\documentclass[useAMS,usenatbib,usegraphicx]{mn2e}  

\usepackage{graphicx,breqn}
\DeclareGraphicsExtensions{.pdf,.eps,.jpg,.ps,.png}
\usepackage{epsfig}
\usepackage{color}
\usepackage{url}
\usepackage{threeparttable}
\usepackage{ amssymb }
\usepackage{url}
\usepackage{subfig}

\newcommand{\bSigma}{\mbox{\boldmath{$\Sigma$} }}
\newcommand{\bTheta}{\mbox{\boldmath{$\Theta$}}}

\newcommand{\bG}{\mbox{\boldmath{$G$}}}

\begin{document}

\title[Bayesian Analysis of the ACS Survey of Galactic Globular Clusters]{The ACS Survey of Galactic Globular Clusters XIV: Bayesian Single-Population Analysis of 69 Globular Clusters}

\author[Wagner-Kaiser et al.]{R. Wagner-Kaiser$^{1}$, A. Sarajedini$^{1}$, T. von Hippel$^{2, 3}$, D. C. Stenning$^{4}$, D. A. van Dyk$^{5}$, 
\newauthor{E. Jeffery$^{6}$, E. Robinson$^{7}$, N. Stein$^{8}$, J. Anderson$^{9}$, W. H. Jefferys$^{10, 11}$} \\
$^1$Bryant Space Center, University of Florida, Gainesville, FL, USA \\
$^2$Center for Space and Atmospheric Research, Embry-Riddle Aeronautical University, Daytona Beach, FL, USA \\
$^3$Max Planck Institute for Astronomy, K\"{o}nigstuhl 17, 69117 Heidelberg, Germany \\
$^4$Sorbonne Universit\'{e}s, UPMC-CNRS, UMR 7095, Institut d'Astrophysique de Paris, F-75014 Paris, France \\
$^5$Imperial College London, London, UK \\
$^6$Brigham Young University, Provo, UT, USA \\
$^7$Argiope Technical Solutions, FL , USA\\
$^8$The Wharton School, University of Pennsylvania, Philadelphia, PA, USA \\
$^9$Space Telescope Science Institute, Baltimore, MD, USA \\
$^{10}$University of Texas, Austin, TX, USA \\
$^{11}$University of Vermont, Burlington, VT, USA \\ }
\date{}

\pagerange{\pageref{firstpage}--\pageref{lastpage}} \pubyear{2014}

\maketitle

\label{firstpage}


\begin{abstract}
We use Hubble Space Telescope (HST) imaging from the ACS Treasury Survey to determine fits for single population isochrones of 69 Galactic globular clusters. Using robust Bayesian analysis techniques, we simultaneously determine ages, distances, absorptions, and helium values for each cluster under the scenario of a ``single" stellar population on model grids with solar ratio heavy element abundances. The set of cluster parameters is determined in a consistent and reproducible manner for all clusters using the Bayesian analysis suite BASE-9. Our results are used to re-visit the age-metallicity relation. We find correlations with helium and several other parameters such as metallicity, binary fraction, and proxies for cluster mass. The helium abundances of the clusters are also considered in the context of CNO abundances and the multiple population scenario.


\end{abstract}

\vspace{2pc}


\section{Introduction}\label{Intro}
Although it is clear that most, if not all, globular clusters harbor two or more populations of stars, these multiple populations do not obviously affect the color-magnitude diagram (CMD) at the visual wavelengths. This is true for many clusters even with CMDs of high photometric quality, such as those obtained from HST imaging. After decades, the visual CMD is well-studied and well-modeled theoretically. Multiple population scenarios agree on a mutual distance and absorption for all populations of stars within the cluster, and most suggest little to no spread in the iron abundances for most clusters. Internal differences in age are also expected to be small for the majority of clusters. Therefore, value remains in using the visual filters F606W and F814W to determine single population CMD-based estimates of various cluster parameters, especially to obtain relative ages, chemical composition, and distances to a high degree of accuracy and precision. This will continue to be true until theoretical models at ultraviolet wavelengths are able to model clusters with similar quality as the visual wavelengths.

For decades, CMDs have been fit with isochrones ``by eye". Occasionally, various numerical techniques (including Bayesian approaches) have been implemented to attempt a determination of the best isochrone (\citealt{Andreuzzi:2011, Jorgensen:2005, Naylor:2006, Hernandez:2008, Janes:2013, Valls-Gabaud:2014}). However, uncertainties in distances, reddenings, and metallicities continue to complicate isochrone fitting attempts, making absolute age measurements of Galactic globular clusters (GGCs) difficult. Many previous studies have focused on comparing the relative ages, which are still capable of increasing our understanding of the formation and evolution of the Galaxy. Studies often focus on measuring differences in color or magnitude from key points in the CMD, such as the main sequence turnoff, red giant branch, and zero age horizontal branch (\citealt{Gratton:1985, Sarajedini:1989, VandenBerg:2000}). An alternative method by \cite{Marin-Franch:2009} used main sequence fitting of deep photometry from Hubble Space Telescope (HST) Advanced Camera for Surveys (ACS) to determine relative ages and distances.

In this paper, we focus on Bayesian analysis, which incorporates many of the above insights as well as a broad range of what we know about globular clusters. We employ principled probability-based methods that embed physics-based models into an overall statistical model that accounts for the complexity of the astronomical sources, thereby leading to greater reliability in practice (\citealt{van-Dyk:2009, De-Gennaro:2009, Stein:2013}). With the computational power available in the modern era, implementing a Bayesian technique is preferable. Besides being more objective than the chi-by-eye approach, Bayesian methods allow simultaneous fitting of multiple characteristics of a cluster along with principled measures of uncertainty (\citealt{van-Dyk:2009, Jeffery:2016}). Bayesian methods also can produce the joint posterior distribution for two or more parameters, allowing us to fully characterize complex non-linear correlations among the fitted parameters. We use a large grid of theoretical models and simultaneous fitting of multiple filters and multiple parameters to obtain greater precision compared to traditional methods. Additionally, unlike standard point estimates and error bars, Bayesian methods allow us to recover the full posterior distribution for each parameter. Together, this makes Bayesian analysis a powerful tool for examining globular clusters and for accurately and precisely determining their characteristics.

With well-determined ages and metallicities derived using Bayesian methods, we can re-visit the age-metallicity relation. The grouping of globular clusters into two groups in the age-metallicity relation is often cited as evidence for the Milky Way halo's two-part formation. A group of old clusters appears to have no dependence on metallicity, while the remaining group of clusters are old and metal-poor, then become younger at higher metallicities (\citealt{Sarajedini:1989, Chaboyer:1992, Richer:1996, Rosenberg:1999, Dotter:2010, Dotter:2011, Leaman:2013}). The older clusters likely formed during a rapid initial assembling of the inner halo, while the other clusters were possibly accreted from the cannibalization of dwarf galaxies over time to the outer halo.

Our Bayesian analysis provides estimates of the overall helium abundances of the clusters, and we use this to explore possible relationships among helium enrichment and other global cluster parameters. Helium is tightly tied to the multiple population scenario in globular clusters (e.g.: \citealt{Bedin:2004, Milone:2009, Gratton:2012, Milone:2012, Piotto:2015}) and we investigate the possible connections herein.

The paper is organized as follows: Section \ref{Data} discusses the data and Section \ref{Methods} provides an overview of the Bayesian software suite. Section \ref{Results} presents the results of the Bayesian fitting, compares our results to recent studies, and explores the correlations we find for the clusters. In Section \ref{Discussion} we discuss the possible implications of our results. We present our conclusions in Section \ref{Conclusion}.


\section{Data}\label{Data}

The data used in this work are primarily from the ACS Globular Cluster Treasury program (GO Cycle 14 Proposal 10775; \citealt{Sarajedini:2007}). This program observed 65 Galactic globular clusters over 132 orbits and produced a consistent set of deep photometry in the HST F606W and F814W filters. Additional data of 6 clusters were obtained from GO-11586 (PI: Dotter) obtained during Cycle 17 (see \citealt{Dotter:2010}). We employ a Bayesian analysis to characterize 69 of the 71 clusters. Palomar 2 is left out of our analysis due to high differential reddening and E3 is removed from the sample due to lack of red giant branch (RGB) stars.

The ACS Globular Cluster Treasury program has provided observations of several hundred thousand stars. To help BASE-9 perform effectively, we make modest quality cuts on the photometry for the clusters before randomly selecting a subsample of stars. To rid the data of any stars with poorly determined photometry, we remove stars for which both filters fall into the outer 5\% tail of the photometric error distributions. Additionally, we remove stars in the outer 2.5\% tails of the distributions of X and Y pixel location errors from frame to frame, as high pixel location errors may indicate non-cluster members. The exceptions to this are the clusters NGC 5986, NGC 6397, and NGC 6779. For these clusters, we ignore the location error constraint as it removes the majority of bright stars above the main sequence turn-off that were observed in the short exposure images. However, we still remove stars in the outer 5\% tail of the photometric error distributions for both filters.

With the cleaned photometry, we randomly select a subsample of $\lesssim$ 3000 stars, with half above the main sequence turnoff point (MSTOP) of the cluster and half below the MSTOP. If there are fewer than 1500 stars above the MSTOP, we match the number of stars above and below the MSTOP. This procedure is adopted to ensure a reasonable sample of stars on the sub-giant and red-giant branches, while balancing their contribution with MS stars, without the computational cost of running 10,000 or more stars per cluster.

\section{Methods}\label{Methods}

\subsection{Bayesian Framework}

Our software suite, known as BASE-9, is a tool for fitting and characterizing observations of open clusters (\citealt{von-Hippel:2006, De-Gennaro:2009, Jeffery:2011, Stein:2013, Hills:2015}), globular clusters (\citealt{Stenning:2016, Wagner-Kaiser:2016, Wagner-Kaiser:2016c}), and even individual stars (\citealt{OMalley:2013}). BASE-9 provides reproducible and precise fits to stellar clusters, determines the cluster-wide parameters of age, distance, metallicity, absorption, and helium fraction, as well as individual stellar parameters of membership, binarity, and mass.

Whether using a numerical method or chi-by-eye, fitting isochrones to stellar clusters often relies on previous studies to assume fixed values of one or several parameters (e.g.: a study interested in age may use [Fe/H] or distance determinations from previous work, or may assume helium abundances). BASE-9 allows us to fix one or more parameters in this way, to use these externally-derived parameters when specifying prior distributions, or, if the data are rich enough, to fit the cluster's parameters simultaneously. Our results presented here also provide a baseline measurement to compare to multiple population studies using similar Bayesian methods (\citealt{Stenning:2016, Wagner-Kaiser:2016}).

BASE-9 uses adaptive Metropolis (AM) Markov chain Monte Carlo (MCMC) to estimate model parameters and map their full posterior distributions (\citealt{Stenning:2016}). It does so by sampling from the joint posterior distribution of distance, metallicity, absorption, age, and helium of a cluster as well as individual parameters of membership, binarity, and mass. In our analyses we fix two of these parameters: binarity and metallicity (as discussed below) and marginalize over the individual stellar parameters.. The adaptive MCMC sampler is designed to explore the joint posterior distribution and use the empirical variance-covariance matrix of past MCMC draws to optimize the proposal distribution. One advantage of the AM algorithm is that the user need only provide starting values for the chain and adaptation is done automatically, such that additional tuning is generally unnecessary. After an initial burn-in period (typically 1000 iterations), we run the AM MCMC chain for at least 10,000 additional iterations. We use one chain per cluster due to computational limitations. Additional details regarding our statistical computation methods and implementation of the adaptive Metropolis algorithm are provided in \cite{Stenning:2016}. 

We assume a fixed value of 0 for $R_i$, thus not including binaries in our model due to computational limitations. Globular clusters tend to have low binary fractions, thus we do not expect that treating stars as singletons should have a significant effect on the final results (see Section \ref{HeliumRelations} for further discussion). Future development of the BASE-9 software will enable the inclusion of binaries in large stellar clusters at reasonable computational cost.

Metallicity values are assumed for our clusters from the spectroscopic measurements in the \cite{Harris:2010} catalog. Because the theoretical models have a strong correlation between metallicity and helium, we cannot gain enough leverage to constrain both simultaneously with BASE-9 using only F606W and F814W. However, fixing metallicity at a reasonable value obtained from the literature allows us to gain precise determinations of helium abundances for a large sample of clusters.

We define a statistical model based on a hierarchy of properties belonging to individual stars or the entire cluster (\citealt{De-Gennaro:2009, van-Dyk:2009, Stein:2013}). The individual parameters are the zero-age main sequence masses of the stars ($M_i$; additional individual parameters are defined shortly). The cluster level parameters are age (base-10 logarithm, $\theta_{\rm age}$), metallicity ($\theta_{[Fe/H]}$), distance modulus ($\theta_{m-M_{V}}$), absorption ($\theta_{A_V}$), and helium fraction ($\phi_{Y}$). We represent the observed photometric data as $X_{ij}$, for the ith star in the jth filter (for $N$ stars and $n$ filters), as described by \cite{De-Gennaro:2009}, \cite{van-Dyk:2009}, \cite{Stein:2013}, and \cite{Stenning:2016}. The known Gaussian measurement errors are in the variance-covariance matrix $\mathbf{\Sigma}_i$ for each star $i$.

For a given set of proposed individual and cluster level parameters, we predict photometry for each of $N$ stars. The function $\bG$ is the stellar evolution model that uses $M_i$, $\Theta$, and $\phi_Y$ to predict an $n$-length vector of magnitudes for 1, $\ldots$, $N$ stars (where $\bTheta$ = ($\theta_{\rm age}$, $\theta_{[Fe/H]}$, $\theta_{m-M_{V}}$, $\theta_{A_V}$), represented as $\mbox{\boldmath{$\mu$}}_i$ = $\bG$($M_i$, $\bTheta$, $\phi_Y$). Although for our purposes helium is a cluster-level parameter, it is treated separately because it is allowed to vary within the cluster in other contexts (\citealt{Stenning:2016, Wagner-Kaiser:2016}). As in \cite{van-Dyk:2009}, binaries can be included in the model by treating each star as a binary system and fitting the component masses. The ratio of the companion mass relative to the primary mass is referred to as $R_i$. 

We represent the predicted magnitude for star $i$ as

\begin{equation}\label{eq:PredMag}
\mathbf{\mbox{\boldmath{$\mu$}}_i} = -2.5 \log_{10}(10^{-\bG(M_i, \bTheta, \phi_Y)/2.5} 
\end{equation}

\noindent where $\mbox{\boldmath{$\mu$}}_i$ is the vector of predicted magnitudes. We also account for the possibility that each star could belong to the Milky Way field population rather than to the cluster itself with a two-component finite mixture distribution (\citealt{van-Dyk:2009}). As such, we define indicator variables, $Z_i$ for $i$=1, ..., $N$, that take the value of 0 for a field star and 1 for a cluster star. We provide a brief description of the main components of the statistical model in Table \ref{tab:likelihood} for reference.

The likelihood function for a single population of stars is
\begin{dmath}\label{eq:LikelihoodSing}
L(\mathbf{M},\mathbf{R},\mathbf{Z},\mathbf{\Theta},\phi_Y \lvert \mathbf{X},\mathbf{\Sigma}) = \prod_{i=1}^{N}\left [ Z_i \times \frac{1}{\sqrt{(2\pi)^n \left  \lvert  \mathbf{\Sigma}_i \right \lvert} }  exp\left (-\frac{1}{2}(\mathbf{X}_i - \mbox{\boldmath{$\mu$}}_i)^\top \times  \mathbf{\Sigma}_i^{-1}(\mathbf{X}_i-\mbox{\boldmath{$\mu$}}_i)   \right ) + (1-Z_i) \times P(\mathbf{X}_i \lvert Z_i=0 ) \right ] .
\end{dmath}

\noindent Equation \ref{eq:LikelihoodSing} may be rewritten as
\begin{dmath}\label{eq:LikelihoodSing2}
L(\mathbf{M},\mathbf{R},\mathbf{Z},\mathbf{\Theta},\phi_Y \lvert \mathbf{X},\mathbf{\Sigma}) = \prod_{i=1}^{N}\left [ Z_i \times P(\mathbf{X}_i \lvert \mathbf{\Sigma}_i, M_i, R_i, \mathbf{\Theta},\phi_Y,Z_i=1)  +(1-Z{i})\times P(\mathbf{X}_i \lvert Z_i=0) \right ] ,
\end{dmath}

\noindent where \textbf{M}  = ($M_1$, ..., $M_N$),  \textbf{R} = ($R_1$, ..., $R_N$), \textbf{Z} = ($\textbf{Z}_1$, ..., $\textbf{Z}_N$), \textbf{X} = ($\textbf{X}_1$, ..., $\textbf{X}_N$), and $\bSigma$ = ($\mathbf{\Sigma}_1$, ..., $\mathbf{\Sigma}_N$). The conditional distribution of observed photometric magnitudes for a star belonging to the cluster is represented by P(\textbf{X$_i$} $\lvert$ $\mathbf{\Sigma}_i$, $M_i$, $R_i$, \textbf{$\Theta$}, $\phi_Y$, $Z_i$=1). Essentially, the first term inside the bracket in Equation \ref{eq:LikelihoodSing2} models the star as if it were a cluster star ($Z_i$=1), while the second term models it as belonging to the field star population.

\begin{table}
\caption{BASE-9 Model} \label{tab:likelihood}
    \begin{tabular}{@{}cl@{}}
\hline {{Notation Symbol}} & {{Description}}  \\ \hline
$M_i$			&	Primary initial stellar mass	 \\
$R_i$			&	Ratio of secondary to primary initial stellar masses 	 \\
$Z_i$			&	Cluster membership indicator  \\
\textbf{$\Theta$}	&	Cluster-level parameters  \\
				&	($\theta_{\rm age}$, $\theta_{[Fe/H]}$, $\theta_{m-M_{V}}$, $\theta_{A_V}$)	 \\
\textbf{$\Phi_{Y}$}	&	Helium abundance of the cluster	\\
\textbf{X$_i$}			&	Photometry for $N$ stars in $n$ filters	\\
$\mathbf{\Sigma}_i$	&	Photometric uncertainties for $N$ stars in $n$ filters	\\  \hline
\end{tabular}
\end{table}

In this way, photometry is predicted at each iteration for a proposed set of stellar and cluster parameters. Over many iterations, we use BASE-9 to explore the parameter space of age, distance, absorption, and helium. The estimates we report are posterior medians of these parameters, from which we can generate a single isochrone for each cluster. Parameters that are not of interest to this study are either marginalized over or fixed, as discussed above.
The user inputs to BASE-9 include the photometry and photometric uncertainty for each star in each filter, prior distributions, and starting values for each cluster parameter. The Dartmouth Stellar Evolution Database theoretical isochrones (\citealt{Dotter:2008}\footnote{\url{http://stellar.dartmouth.edu/}}) presents a multi-dimensional space for the MCMC sampler to step through. We have generated a grid of isochrones to cover a range of ages from 1 to 15 Gyr, a range of [Fe/H] metallicity from --2.5 to 0.5 dex, and a range of helium values from $\sim$0.23 to 0.4. This DSED grid has been generated at a [$\alpha$/Fe] value of 0.0 (solar) and we assume an R$_{V}$=3.1 reddening law (\citealt{Cardelli:1989}) during the analysis.

\subsection{Prior Distributions}
 
Because we perform a Bayesian analysis, we must specify prior distributions that summarize our knowledge regarding the model parameters before considering the current data. Previous studies are used to generate Gaussian prior distributions for distance and absorption for each cluster (\citealt{Harris:2010}). Generally speaking these priors are centered on published estimates but we use conservative prior dispersions (e.g., relatively large prior standard deviations) to reduce the influence of the priors on our final estimates. For the distance prior, we choose $\sigma$(m$-$M) = 0.05 mag. For absorption, we conservatively use $\sigma$($A_{\rm V}$) = (1/3) $A_{\rm V}$, using a Gaussian truncated at zero. Because there is no useful prior information on helium or absolute age, a uniform prior on helium of 0.15 to 0.45 is assumed and a uniform prior on age of 1 to 13.5 Gyr. We impose an upper limit of 13.5 Gyr, because although the models extend to 15 Gyr, we do not expect GGCs to be older than the epoch of galaxy formation (\citealt{Pilipenko:2013, Oesch:2016}). 

For helium, Y=0.24 is adopted as a starting point for all clusters. Ages from \citet{Dotter:2010} and \citet{Roediger:2014} are used as starting values for the age. For the clusters not in either of these studies, we adopt a starting point of 12 Gyr for the age. A \cite{Miller:1979} initial mass function defines the prior distribution for initial stellar mass, over the range of 0.1 to 8 solar masses (\citealt{von-Hippel:2006}, \citealt{De-Gennaro:2009}, \citealt{Stenning:2016}).

The cluster membership represents whether a star belongs to the cluster, as opposed to the field star population. As the cleaned photometry discussed above is expected to be largely free of field star contamination, we adopt a incidence rate of field stars of 5\%, as we expect different values of the membership probability in the corresponding range of 0.9 to 1.0 produce consistent results (\citealt{Stenning:2016}). Blue straggler stars are not included in our model. As previously mentioned, because the primary interest is in helium, which is strongly correlated with [Fe/H], the spectroscopic metallicities from \cite{Harris:2010} are used to set the metallicity to previously measured values. 

\begin{figure*}
\centering
\includegraphics[width=0.9\textwidth]{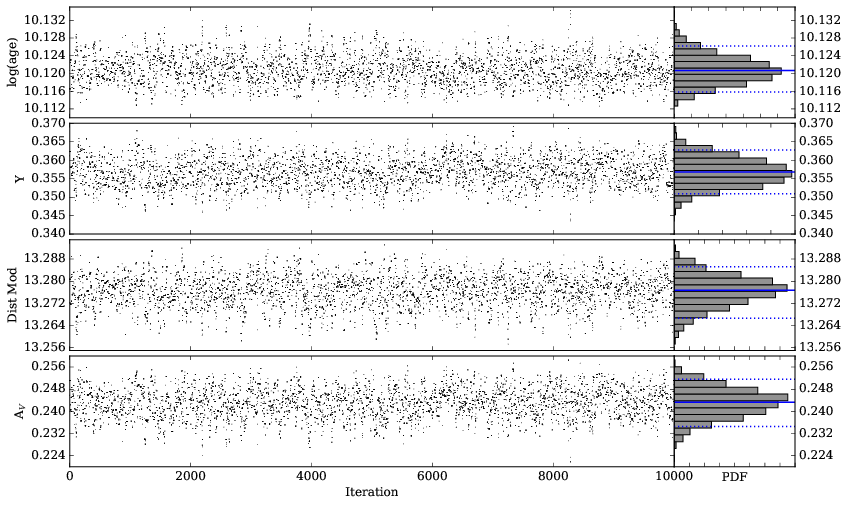}
\caption{The posterior draws of NGC 6752 for 10,000 iterations. From top to bottom, the MCMC chain for log(age), helium, distance modulus, and absorption. On the right, the posterior distribution is shown, with the solid line indicating the median and the dotted lines marking the 90\% central Bayesian credible intervals.}
\label{samplinghist}
\end{figure*}

\begin{figure*}
\centering
\includegraphics[width=0.9\textwidth]{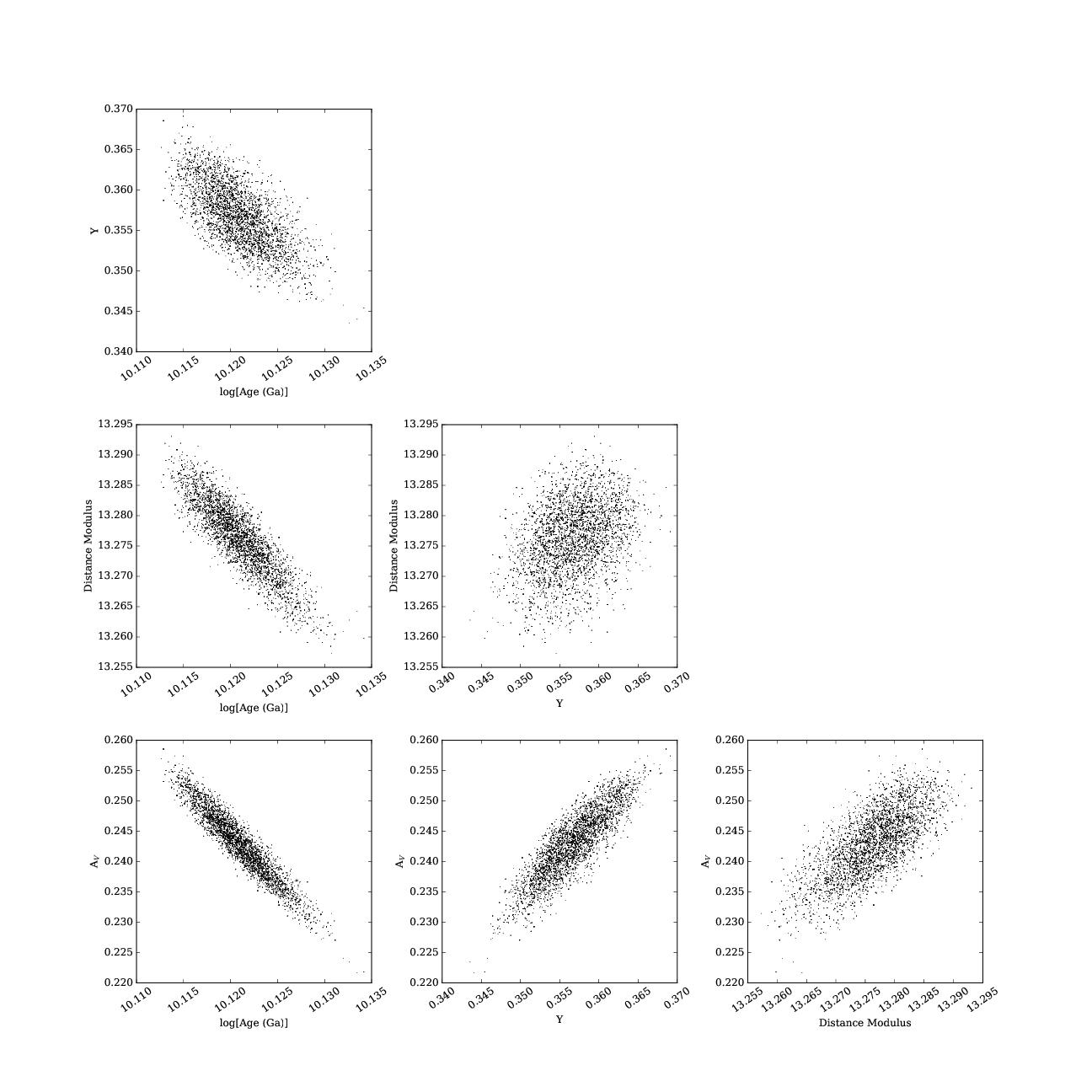}
\caption{A 2-dimensional comparison of the posterior draws for NGC 6752 for the parameters of log(age), helium, distance modulus, and absorption.}
\label{2Dcorner}
\end{figure*}

The clusters are listed in Table \ref{clusterlist} with their adopted prior distributions and starting values. We provide an example of the posterior draws from BASE-9 as well as a 2-D correlation plot in Figures \ref{samplinghist} and \ref{2Dcorner} for NGC 6752.

\begin{table*}
\caption{ACS Treasury Galactic Globular Clusters$^a$ (Full Table Available Online)}
\centering
\begin{threeparttable}[b]
    \begin{tabular}{@{}|c|c|c|c|c|@{}}
    \hline
     \textbf{Cluster} & \multicolumn{2}{c}{\textbf{Prior Distribution}}& \textbf{Assumed Value} & \textbf{Starting Value} \\ \cline{2-3}
 \textbf{Name} & \textbf{Distance Modulus}  & \textbf{A$_{V}$} & \textbf{[Fe/H]}    & \textbf{Age (Gyr)} \\  
\hline
Arp2 	 & 17.59 $\pm$0.05 	 & 0.31 $\pm$ 0.103 	 & -1.75 	 & 13.0 $^b$ \\
IC4499 	 & 17.08 $\pm$0.05 	 & 0.713 $\pm$ 0.238 	 & -1.53 	 & 12.0 $^b$ \\
Lynga7 	 & 16.78 $\pm$0.05 	 & 2.263 $\pm$ 0.754 	 & -1.01 	 & 12.5 $^b$ \\
NGC0104 	 & 13.37 $\pm$0.05 	 & 0.124 $\pm$ 0.041 	 & -0.72 	 & 12.75 $^b$ \\
NGC0288 	 & 14.84 $\pm$0.05 	 & 0.093 $\pm$ 0.031 	 & -1.32 	 & 12.5 $^b$ \\
NGC0362 	 & 14.83 $\pm$0.05 	 & 0.155 $\pm$ 0.052 	 & -1.26 	 & 11.5 $^b$ \\
NGC1261 	 & 16.09 $\pm$0.05 	 & 0.031 $\pm$ 0.010 	 & -1.27 	 & 11.5 $^b$ \\
NGC1851 	 & 15.47 $\pm$0.05 	 & 0.062 $\pm$ 0.021 	 & -1.18 	 & 10.0 $^c$ \\
NGC2298 	 & 15.6 $\pm$0.05 	 & 0.434 $\pm$ 0.145 	 & -1.92 	 & 13.0 $^b$ \\
NGC2808 	 & 15.59 $\pm$0.05 	 & 0.682 $\pm$ 0.227 	 & -1.14 	 & 10.9 $^c$ \\
\hline
    \end{tabular}
\begin{tablenotes}[b]
	\item $^a$ Parameters from \protect\citet{Harris:2010}, unless otherwise noted.
	\item $^b$ From \protect\citet{Dotter:2010}.
	\item $^c$ From \protect\citet{Roediger:2014}.
	\item $^d$ No parameters from either study; 12 Gyr is used as starting age.
\end{tablenotes}
\end{threeparttable}
\label{clusterlist}
\end{table*}

\section{Results}\label{Results}

In Table \ref{resTable}, we provide the results from our Bayesian analyses. The values are the median of the posterior distributions for each parameter and the errors reflect 90\% central Bayesian credible intervals from the BASE-9 posterior distribution. These errors represent statistical uncertainty, assuming the validity of the models (discussed further in Section \ref{Discussion}). We use the medians and credible intervals in subsequent analyses and tables, with the quoted errors representing these statistical uncertainties.

Figure \ref{CMDS1} presents CMDs for each cluster. All ACS photometry are shown as gray points, and the subsamples of stars used in the BASE-9 analysis are are shown in black. The cyan isochrones are generated using the median of the posterior distributions for the parameters, as presented in Table \ref{resTable}.

Many clusters in our sample have light element abundances at non-solar ratios. \cite{Dotter:2010} classifies clusters with abundances of [$\alpha$/Fe] = 0.0, 0.2, and 0.4. However, DSED models are only available with varying helium for the two [alpha/Fe] endpoints, at [$\alpha$/Fe] = 0.0 and 0.4. For the clusters identified as [$\alpha$/Fe] = 0.4 in \cite{Dotter:2010}, we run them with a grid of [$\alpha$/Fe] = 0.4 instead of [$\alpha$/Fe] = 0.0. We present these results in Table \ref{resTable_a4}, and show the resulting isochrone in red in Figure \ref{CMDS1}. Aside from NGC 288, the [$\alpha$/Fe] = 0.4 model grid does not appear to result in visually satisfying fits.

\renewcommand{\arraystretch}{2}
\begin{table*}
\caption{Bayesian Single Population Analysis of 69 Galactic Globular Clusters (Full Table Available Online)}
\label{resTable}
\centering
\begin{threeparttable}[b]
    \begin{tabular}{@{}cccccccc@{}}
    \hline
 \textbf{Cluster} & \textbf{[Fe/H]} & \textbf{Distance} & \textbf{A$_{V}$}  & \textbf{Y}	& \textbf{log(Age)}	& \textbf{Age (Gyr)}	& \textbf{Relative Age (Gyr)}$^a$  \\
 \hline
Arp2 	 & -1.750 	 & 17.613$^{+0.012}_{-0.013}$ 	 & 0.350$^{+0.006}_{-0.006}$ 	 & 0.300$^{+0.009}_{-0.009}$ 	 & 10.130$^{+0.001}_{-0.003}$ 	 & 13.476$^{+0.022}_{-0.083}$ 	 & 0.476 \\
IC4499 	 & -1.530 	 & 17.006$^{+0.006}_{-0.006}$ 	 & 0.741$^{+0.003}_{-0.004}$ 	 & 0.329$^{+0.001}_{-0.002}$ 	 & 10.102$^{+0.003}_{-0.003}$ 	 & 12.661$^{+0.077}_{-0.078}$ 	 & 0.661 \\
Lynga7 	 & -1.010 	 & 16.781$^{+0.015}_{-0.016}$ 	 & 2.483$^{+0.014}_{-0.013}$ 	 & 0.291$^{+0.012}_{-0.011}$ 	 & 10.097$^{+0.011}_{-0.011}$ 	 & 12.513$^{+0.321}_{-0.319}$ 	 & 0.013 \\
NGC0104 	 & -0.720 	 & 13.378$^{+0.004}_{-0.004}$ 	 & 0.105$^{+0.002}_{-0.002}$ 	 & 0.271$^{+0.003}_{-0.003}$ 	 & 10.130$^{+0.000}_{-0.001}$ 	 & 13.494$^{+0.006}_{-0.022}$ 	 & 0.744 \\
NGC0288 	 & -1.320 	 & 14.948$^{+0.009}_{-0.009}$ 	 & 0.084$^{+0.008}_{-0.007}$ 	 & 0.291$^{+0.006}_{-0.006}$ 	 & 10.101$^{+0.005}_{-0.005}$ 	 & 12.629$^{+0.149}_{-0.157}$ 	 & 0.129 \\
NGC0362 	 & -1.260 	 & 14.852$^{+0.006}_{-0.006}$ 	 & 0.117$^{+0.006}_{-0.006}$ 	 & 0.302$^{+0.006}_{-0.005}$ 	 & 10.059$^{+0.004}_{-0.004}$ 	 & 11.457$^{+0.097}_{-0.105}$ 	 & -0.043 \\
NGC1261 	 & -1.270 	 & 16.110$^{+0.005}_{-0.006}$ 	 & 0.070$^{+0.005}_{-0.005}$ 	 & 0.303$^{+0.005}_{-0.004}$ 	 & 10.065$^{+0.003}_{-0.003}$ 	 & 11.606$^{+0.092}_{-0.090}$ 	 & 0.106 \\
NGC1851 	 & -1.180 	 & 15.526$^{+0.006}_{-0.005}$ 	 & 0.157$^{+0.005}_{-0.005}$ 	 & 0.307$^{+0.005}_{-0.004}$ 	 & 10.058$^{+0.004}_{-0.003}$ 	 & 11.433$^{+0.096}_{-0.092}$ 	 & --- \\
NGC2298 	 & -1.920 	 & 15.627$^{+0.012}_{-0.026}$ 	 & 0.760$^{+0.008}_{-0.005}$ 	 & 0.337$^{+0.018}_{-0.008}$ 	 & 10.130$^{+0.000}_{-0.001}$ 	 & 13.493$^{+0.007}_{-0.017}$ 	 & 0.493 \\
NGC2808 	 & -1.140	 & 15.658$^{+0.006}_{-0.007}$ 	 & 0.648$^{+0.003}_{-0.003}$ 	 & 0.329$^{+0.001}_{-0.002}$ 	 & 10.045$^{+0.003}_{-0.003}$ 	 & 11.092$^{+0.080}_{-0.075}$ 	 & --- \\
  \hline
    \end{tabular}
\begin{tablenotes}[b]
	\item $^a$ Relative to \protect\citet{Dotter:2010} and \protect\citet{Dotter:2011}, where applicable.
\end{tablenotes}
\end{threeparttable}
\end{table*}
\renewcommand{\arraystretch}{1}

\renewcommand{\arraystretch}{2}
\begin{table*}
\caption{Bayesian Single Population Analysis of Galactic Globular Clusters with [$\alpha$/Fe]=0.4}
    \label{resTable_a4}
    \centering
\begin{threeparttable}[b]
    \begin{tabular}{@{}ccccccc@{}}
       \hline
 \textbf{Cluster} & \textbf{[Fe/H]} & \textbf{Distance} & \textbf{A$_{V}$}  & \textbf{Y}	& \textbf{log(Age)}	& \textbf{Age (Gyr)}  \\
 \hline
NGC0288 	 & -1.32 	 & 15.011$_{-0.005}^{+0.005}$ 	 & 0.149$_{-0.005}^{-0.005}$ 	 & 0.358$_{-0.006}^{+0.005}$ 	 & 10.13$_{-0.000}^{+0.000}$ 	 & 13.497$_{-0.011}^{+0.003}$ \\
NGC4833 	 & -1.85 	 & 15.187$_{-0.008}^{+0.008}$ 	 & 1.326$_{-0.005}^{-0.007}$ 	 & 0.496$_{-0.007}^{+0.010}$ 	 & 10.13$_{-0.000}^{+0.000}$ 	 & 13.496$_{-0.011}^{+0.004}$ \\
NGC6121 	 & -1.16 	 & 12.851$_{-0.004}^{+0.004}$ 	 & 1.437$_{-0.002}^{-0.003}$ 	 & 0.33$_{-0.001}^{+0.003}$ 	 & 10.13$_{-0.001}^{+0.000}$ 	 & 13.493$_{-0.027}^{+0.007}$ \\
NGC6352 	 & -0.64 	 & 14.524$_{-0.014}^{+0.008}$ 	 & 0.86$_{-0.003}^{-0.005}$ 	 & 0.333$_{-0.004}^{+0.007}$ 	 & 10.13$_{-0.001}^{+0.000}$ 	 & 13.495$_{-0.018}^{+0.005}$ \\
NGC6362 	 & -0.99 	 & 14.749$_{-0.005}^{+0.006}$ 	 & 0.305$_{-0.002}^{-0.002}$ 	 & 0.327$_{-0.004}^{+0.003}$ 	 & 10.13$_{-0.000}^{+0.000}$ 	 & 13.497$_{-0.011}^{+0.003}$ \\
NGC6426 	 & -2.15 	 & 17.696$_{-0.104}^{+0.011}$ 	 & 1.628$_{-0.005}^{-0.055}$ 	 & 0.618$_{-0.008}^{+0.058}$ 	 & 10.124$_{-0.010}^{+0.003}$ 	 & 13.301$_{-0.308}^{+0.093}$ \\
NGC6541 	 & -1.81 	 & 14.937$_{-0.007}^{+0.008}$ 	 & 0.610$_{-0.004}^{-0.005}$ 	 & 0.437$_{-0.007}^{+0.007}$ 	 & 10.13$_{-0.001}^{+0.000}$ 	 & 13.491$_{-0.025}^{+0.008}$ \\
Palomar15 	 & -2.07 	 & 19.604$_{-0.021}^{+0.019}$ 	 & 1.548$_{-0.008}^{-0.007}$ 	 & 0.331$_{-0.004}^{+0.010}$ 	 & 10.13$_{-0.002}^{+0.000}$ 	 & 13.484$_{-0.053}^{+0.012}$ \\
Terzan8 	 & -2.16 	 & 17.756$_{-0.015}^{+0.014}$ 	 & 0.661$_{-0.007}^{-0.008}$ 	 & 0.433$_{-0.012}^{+0.013}$ 	 & 10.13$_{-0.001}^{+0.000}$ 	 & 13.489$_{-0.045}^{+0.011}$ \\
  \hline
    \end{tabular}
\end{threeparttable}
\end{table*}
\renewcommand{\arraystretch}{1}

\begin{figure*}
\centering
\includegraphics[width=0.9\textwidth]{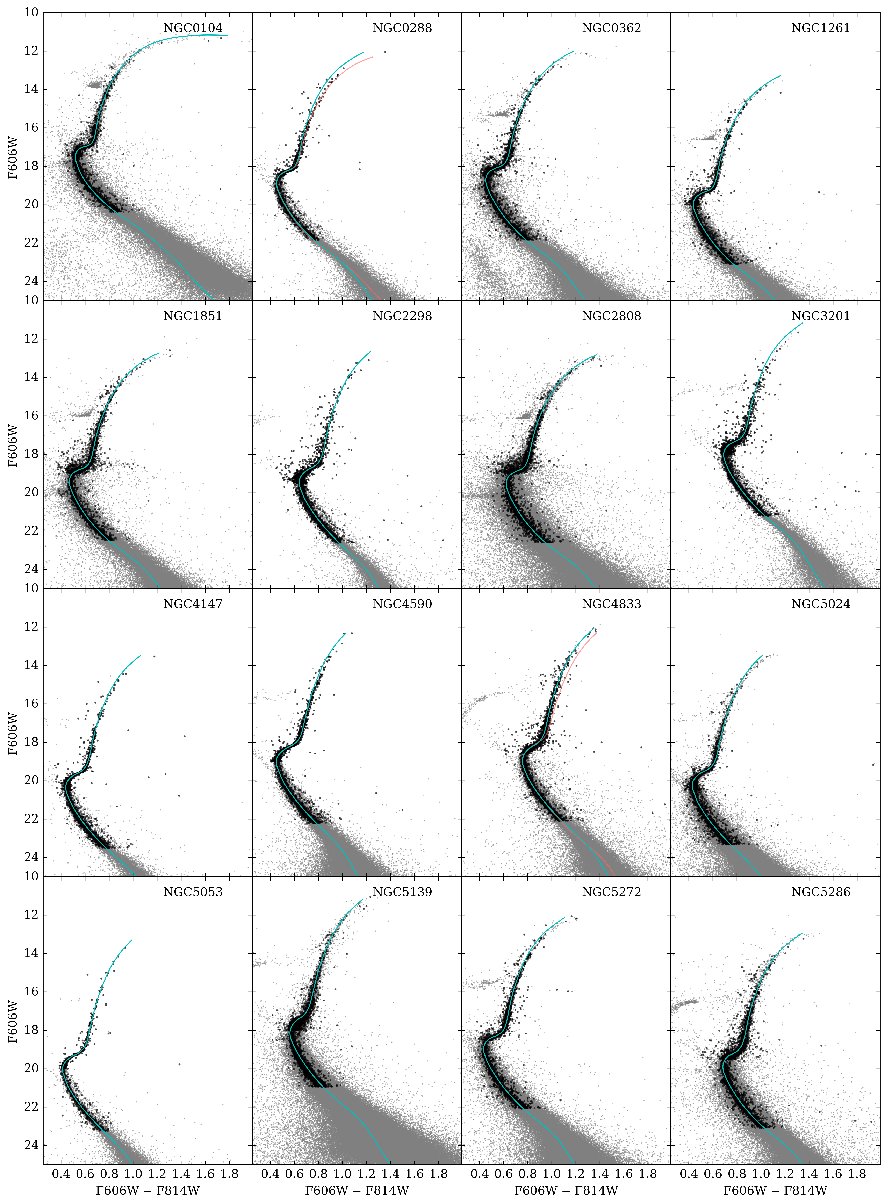}
\caption{Results of single population Bayesian analysis on the globular clusters for the grid of DSED models. Gray points show the published photometry from the ACS Globular Cluster Treasury (\citealt{Sarajedini:2007}). The black points show the subsample of stars for each cluster that are run with BASE-9. The solid, cyan line shows the isochrone generated from the medians of the posterior distributions of the sampled parameters for each cluster (i.e.: age, distance, A$_{V}$, and helium). The red isochrone indicates the same, using the DSED model grid with [$\alpha$/Fe] = 0.4 for some clusters.}
\label{CMDS1}
\end{figure*}

\begin{figure*}
\ContinuedFloat
\centering
\includegraphics[width=0.95\textwidth]{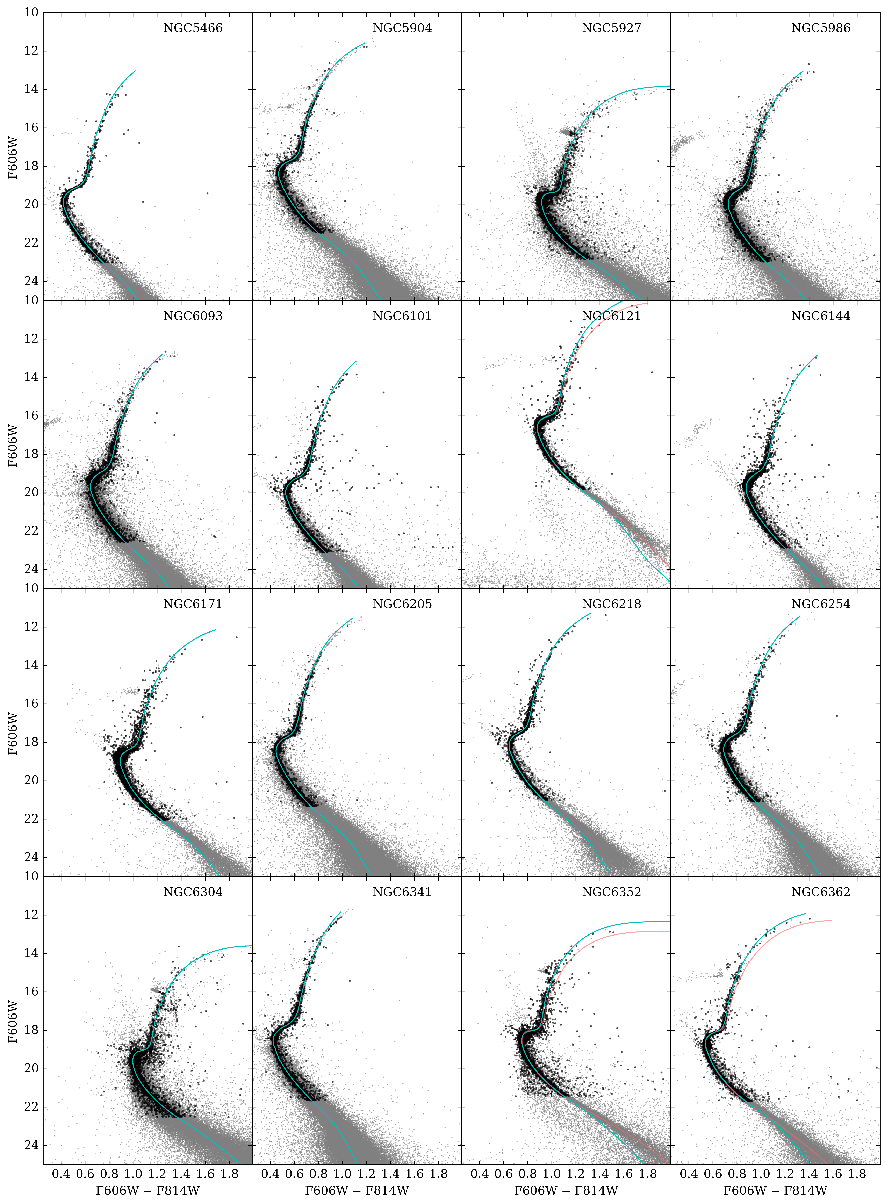}
\caption{Continued, for additional clusters.}
\label{CMDS2}
\end{figure*}

\begin{figure*}
\ContinuedFloat
\centering
\includegraphics[width=0.95\textwidth]{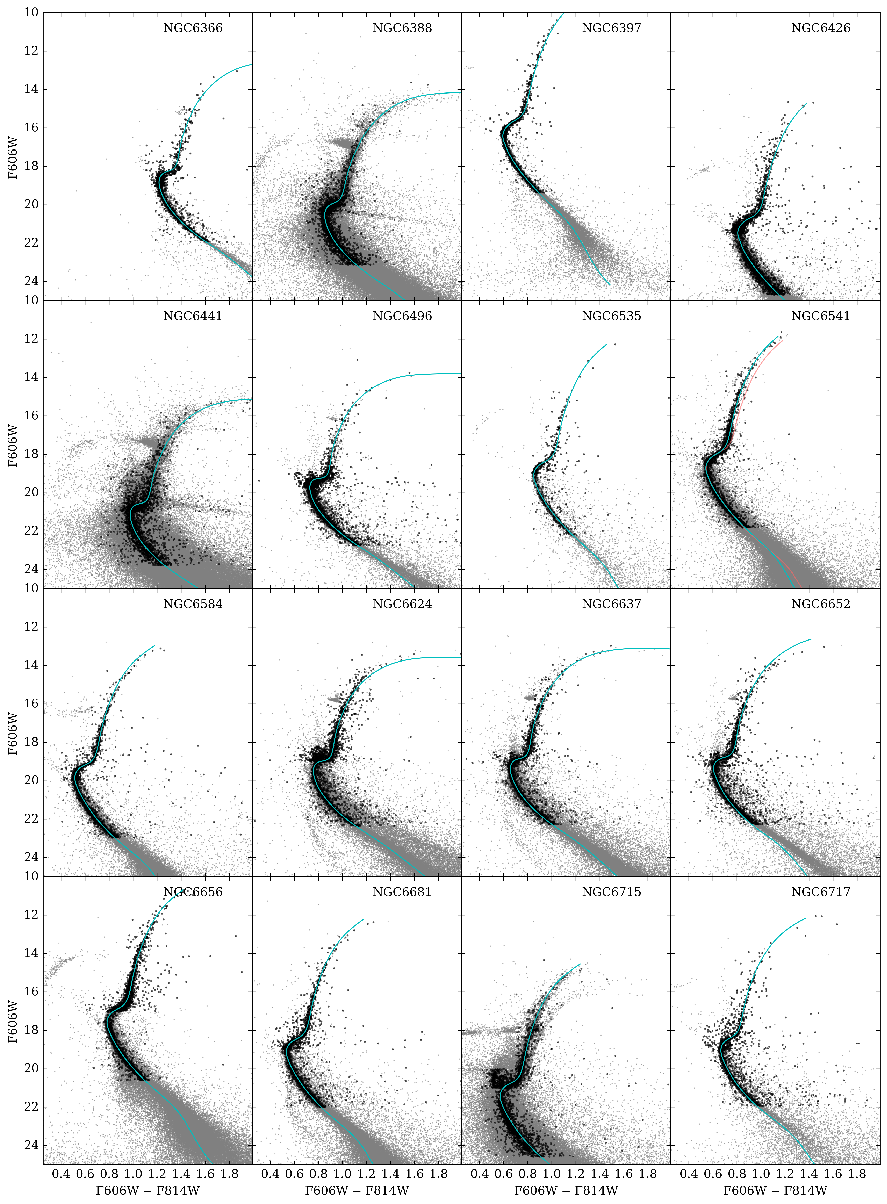}
\caption{Continued, for additional clusters.}
\label{CMDS3}
\end{figure*}

\begin{figure*}
\ContinuedFloat
\centering
\includegraphics[width=0.95\textwidth]{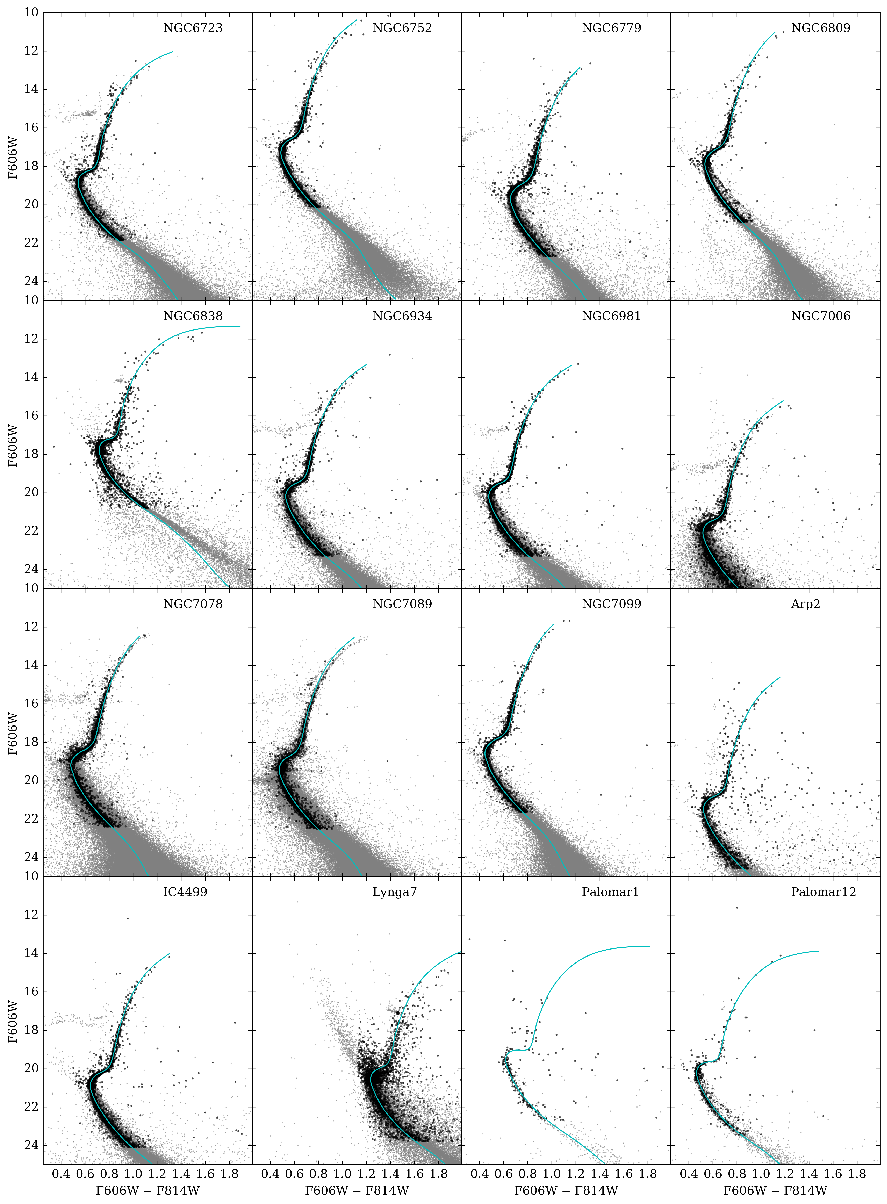}
\caption{Continued, for additional clusters.}
\label{CMDS4}
\end{figure*}

\begin{figure*}
\ContinuedFloat
\centering
\includegraphics[width=0.95\textwidth]{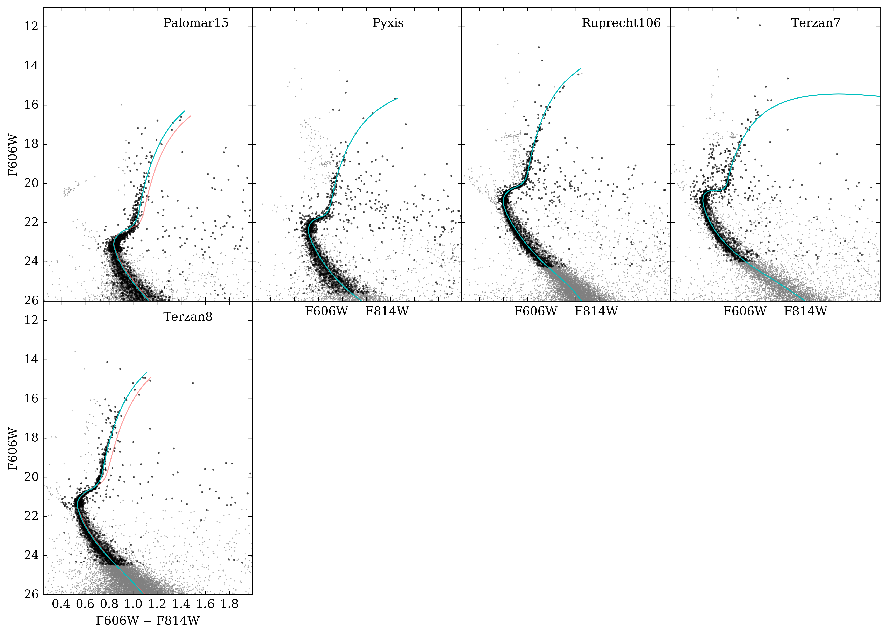}
\caption{Continued, for additional clusters.}
\label{CMDS5}
\end{figure*}


\subsection{Comparisons to Previous Work}\label{Comparison}

In this section, we present a brief comparison of our results to previously published results, specifically those of \citet{Dotter:2010} and \citet{Harris:2010}. Because we allow the helium fraction to vary as opposed to previous studies which fix or assume its value, other parameters change to maintain a best fit of the models to the data, so differences with published studies are not unexpected.

\begin{figure*}
\centering
\includegraphics[width=\textwidth]{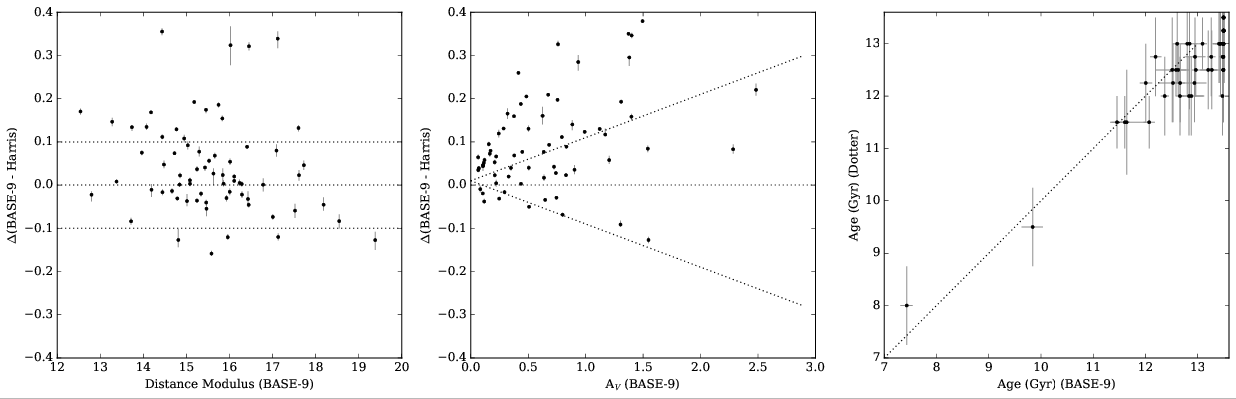}
\caption{Our results compared to that of \protect\citet{Dotter:2010} and \protect\citet{Harris:2010}. The leftmost panel plots the difference between estimated distance moduli from \protect\citet{Harris:2010} and this study vs. our distance modulus. The dotted horizontal lines indicate the $\pm$0.1 errors in distance suggested by the \protect\citet{Harris:2010} catalog. The middle panel repeats the comparison for absorption, with the dotted line delineating the $\pm$0.1*Av errors for absorption from \protect\citet{Harris:2010}. In both these panels, the vertical errors are our 90\% central Bayesian credible intervals for distance and absorption. The rightmost panel shows a comparison between the estimated age from \protect\citet{Dotter:2010} and our median posterior age.}
\label{Comp}
\end{figure*}

In Figure \ref{Comp}, we show a comparison of the median of our posterior distribution for distance and absorption results to the estimates for each cluster from \citet{Harris:2010} in the left and middle panels. We calculate the differences as the \citet{Harris:2010} estimates subtracted from our own. In these panels, the errorbars represent our BASE-9 90\% central Bayesian credible intervals and we indicate the error estimates from \cite{Harris:2010} as dotted lines. Specifically, an error of $\pm$0.1 for distance and $\pm$0.1*Av for absorption. Our distance estimates largely fall within the range of error from of distances from the \cite{Harris:2010} catalog. As one might expect, it appears that for larger values of absorption, the scatter increases. In many cases, we estimate moderately larger absorption than the values from \cite{Harris:2010}.

In the rightmost panel of Figure \ref{Comp} we compare the BASE-9 and \cite{Dotter:2010} estimates of age\footnote{We do not plot $\Delta$(BASE-9 - Dotter) on the vertical axis of the rightmost panel of Figure \ref{Comp} because errors on $\Delta$(BASE-9 - Dotter) are not readily computable. This is because the Dotter et al and the BASE-9 estimates are compiled using the same ACS photometry from \cite{Sarajedini:2007}, leading to correlation between the estimates so that the error on $\Delta$(BASE-9 - Dotter) is not simply the sum (in quadrature) of the errors of the two estimates.}. While the two studies largely agree within the (correlated) errors, it is not unexpected that we should find slightly different results from \citet{Dotter:2010} as we explore variations in helium and not in metallicity. We provide relative ages where available for the clusters with respect to \citet{Dotter:2010} in the final column of Table \ref{resTable}.

\subsection{Age$-$Metallicity Relation}\label{AMR}

\begin{figure}
\includegraphics[width=0.5\textwidth]{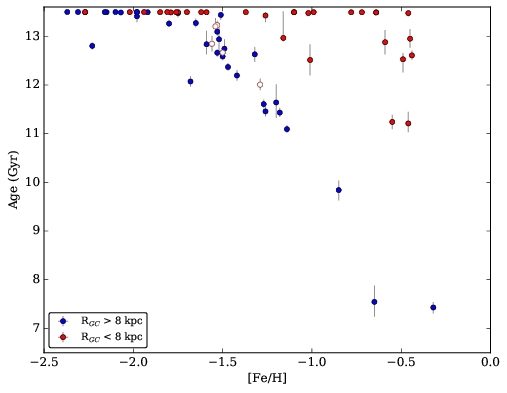}
\caption{The age-metallicity relation for 69 clusters with BASE-9 fits. Red circles indicate the inner halo clusters within 8 kpc, and blue circles indicate the outer halo clusters beyond 8 kpc. The inner halo clusters that appear to follow the trend of the outer halo clusters are indicated with a white circle outlined in red.}
\label{AMRplot}
\end{figure}

We revisit the age-metallicity relationship with our results, as seen in Figure \ref{AMRplot}. The age-metallicity relation (AMR) has long been held as support for a two-phase formation scenario of the Milky Way (\citealt{Chaboyer:1996}, \citealt{Marin-Franch:2009}, \citealt{VandenBerg:2013}). This view of the formation of the Galaxy suggests that the bulk of the inner halo formed early-on along with the metal-poor clusters of the outer halo. Cluster formation then continued mainly in the outer halo, forming more metal-rich and younger clusters.

Consistent with previous studies, we find a group of old clusters with a relatively narrow age range almost exclusively found in the inner 8 kpc of the Galaxy, with a error-weighted mean age of 13.433$\pm$1.005 Gyr (log(age) of 10.124$\pm$0.033). The group of clusters beyond 8 kpc follow a different trend, such that as these clusters become more metal-rich, they also become younger with more significant scatter. For these clusters, we find an error-weighted mean age of 12.320$\pm$1.755 Gyr (log(age) of 10.067$\pm$0.069). We see a clear distinction between the two sequences and consider this strong evidence for the two-part formation history of the Milky Way. The old, inner halo clusters likely formed quickly early on in the history of the Galaxy, while the younger trend of clusters slowly formed or joined the outer halo over time via chaotic accretion. This is also thought to be tied to the horizontal branch morphology--age relationship and the second parameter problem (\citealt{Searle:1978,van-den-Bergh:1965, Sarajedini:1989,Dotter:2010}). However, our results do not show clear trends with helium in the context of the age-metallicity relation.

Many cluster ages are pinned against the age of the Universe, an external prior built into our analysis.  This indicates that the isochrones prefer an older fit. Nonetheless, despite the limitations in the absolute ages of isochrone fitting, we are focusing on relative ages, specifically on distinguishing the median age and age spread of the inner clusters versus the outer clusters.

There are several inner clusters within 8 kpc of the Galactic center that we find along the ``wrong" sequence. One of these is $\omega$ Cen (NGC 5139), which is both the most massive globular cluster and which likely contains the largest number of stellar populations (\citealt{Bedin:2004, Bekki:2006,Bellini:2010,Johnson:2008}). NGC 6584 has a large rotational velocity, making it probable that it does not truly belong to the ``inner" galactic clusters. Instead, it is likely an outer halo cluster that is observed to be passing through the inner halo. It's unclear why the other clusters, NGC 5904, NGC 6254, and NGC 6752 follow the sequence of outer halo clusters rather than that of the inner clusters. We indicate these clusters with a red outline in Figure \ref{AMRplot} to show their locations.

\subsection{Helium}\label{HeliumRelations}

In this section, we present a variety of correlations we find among helium and other cluster parameters, and discuss the possible implications of these findings. 

\begin{figure}
\includegraphics[width=0.5\textwidth]{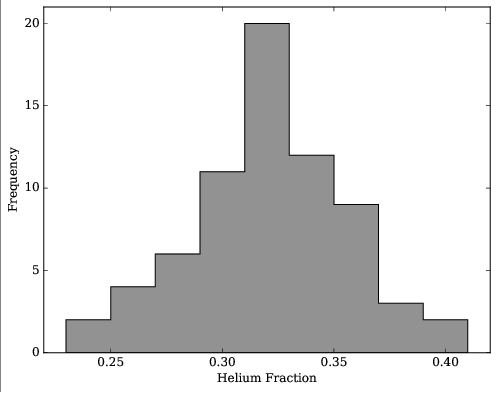}
\caption{The distribution of helium values obtained for the 69 Galactic globular clusters included in this study. Fitted helium values range from 0.235 to 0.409 with a peak at about 0.321.}
\label{Helium_hist}
\end{figure}

The distribution of fitted helium fractions in the clusters is approximately symmetric, as seen in Figure \ref{Helium_hist}, with a total range of about 0.17, from Y = 0.235 (NGC 6652) to Y = 0.409 ($\omega$ Cen). The peak in the distribution is at Y = 0.321$\pm$0.035. Because we are probing the overall (``average") helium abundance of the cluster, we do not expect helium abundances lower than the primordial value. Our lowest estimated helium abundance is 0.235, similar to the lower limit on the primordial value (we list a few recent values in Table \ref{Primordial} for comparison). As almost, if not all, these clusters harbor multiple populations, the second (or third, fourth, etc.) generation stars in each cluster may be more enriched in helium. The helium abundances measured should be population-weighted values (e.g.: a higher measurement of helium could be caused by a higher proportion of later generations of stars, a higher level of helium enrichment, or both). Our helium fraction estimates are also sensitive to our assumption of metallicity abundances, with more metal-rich clusters being more sensitive to the assumed metallicity.

\renewcommand{\arraystretch}{2}
\begin{table}
\centering
    \caption{Primordial Helium Values from Literature}
    \begin{tabular}{@{}cl@{}}
    \hline
 \textbf{Primordial Helium} & \textbf{Reference}   \\
 \hline
0.232$^{+0.044}_{-0.047}$ to 0.293$^{+0.046}_{--0.048}$	 & \citet{Planck-Collaboration:2014} \\
0.254$\pm0.003$			 & \citet{Izotov:2013}		\\
0.278$^{+0.034}_{-0.032}$ 	 & \citet{Hinshaw:2013} 	  \\
0.2485$\pm$0.0002 			 & \citet{Aver:2013}	  \\
0.2477$\pm$0.0029 			 &\citet{Peimbert:2007}	  \\
0.250$\pm$0.006			 &\citet{Salaris:2004} \\
\hline
    \end{tabular}
   \label{Primordial}
\end{table}
\renewcommand{\arraystretch}{1}

\begin{figure}
\includegraphics[width=0.5\textwidth]{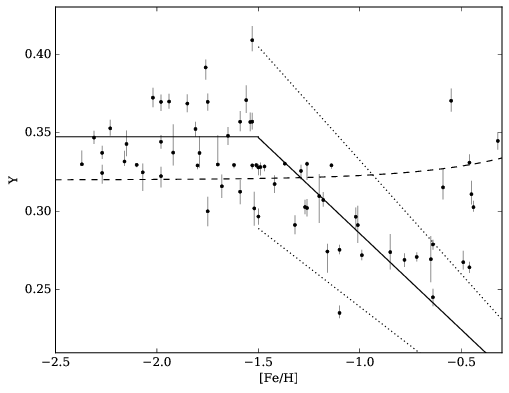}
\caption{BASE-9 fitted helium values plotted against metallicity. We fit a broken relation (indicated by the solid line), with Y = 0.348 for [Fe/H]$\textless$--1.5 and a slope of --0.122$\pm$0.022 and an intercept of 0.165$\pm$0.024 for [Fe/H]$\textgreater$--1.5. We find a correlation of --0.545$\pm$0.001. The dashed line represents the assumption of $\Delta$Y/$\Delta$Z = 1.54 as employed by the \protect\citet{Dotter:2008} models. Dotted lines indicate 1-$\sigma$ errors in both the slope and intercept.}
\label{Y_FeH}
\end{figure}

We find a statistically significant correlation between the helium abundance of a cluster and its metallicity, as seen in Figure \ref{Y_FeH}, with errors given by the 90\% central Bayesian credible intervals from Table \ref{resTable}. We fit a broken line to the data (solid line), with Y = 0.348 for [Fe/H]$\textless$--1.5 and Y = --0.122($\pm$0.022) [Fe/H] + 0.165($\pm0.024$) for [Fe/H]$\textgreater$-1.5. The relationship has an error-weighted Pearson correlation coefficient of --0.545$\pm$0.001. 

Overall, we find that the metal-poor clusters have similar helium abundances. As clusters become more metal rich, they appear to become less helium-rich. However, the clusters NGC 6304, NGC 6366, NGC 6388, NGC 6441, NGC 6624, and Terzan 7 do not follow this trend, instead being well-explained by the model assumption from \cite{Dotter:2008} of $\Delta$Y/$\Delta$Z = 1.54.

The spread in helium for metal-rich clusters could be due to a high percentage of chemically unenriched stars or due to less helium enrichment compared to other clusters. In our study, we are unable to distinguish between the two effects.

\begin{figure}
\includegraphics[width=0.5\textwidth]{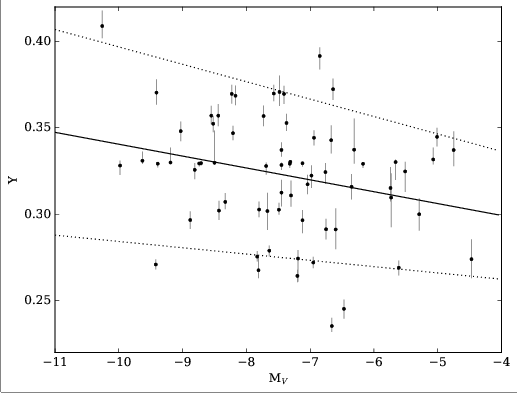}
\caption{The absolute integrated V magnitude of the cluster from \protect\citet{Harris:2010}, which is a proxy for mass, versus helium fraction. These quantities are positively correlated, suggesting that more massive clusters have more enriched helium and light element abundances. The linear regression (solid line) including errors from our helium estimates has a slope of --0.007$\pm$0.003 and an intercept of 0.272$\pm$0.024. Dashed lines indicate 1-$\sigma$ errors in both the slope and intercept.}
\label{Y_Mabs}
\end{figure}

More massive clusters are predicted to attain higher enrichments of helium (\citealt{Milone:2014}, \citealt{Milone:2015}). As shown in Figure \ref{Y_Mabs}, we find that the absolute magnitude of the clusters from \citet{Harris:2010} (a proxy for the mass of the cluster) is likely correlated to the helium of the cluster, such that brighter clusters tend to have marginally higher values of helium. The weighted Pearson correlation coefficient between helium and the absolute magnitude is moderate at --0.261$\pm$0.032, but the trend suggests support for previous claims.

\begin{figure}
\includegraphics[width=0.5\textwidth]{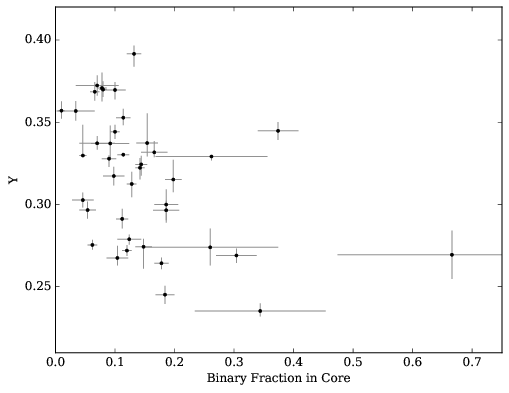}
\caption{The binary fraction cluster cores vs. the cluster helium content. For clusters with a binary fraction less than 20\%, higher abundances of helium appear correlated with a lower fraction of binaries.}
\label{Y_Binaries}
\end{figure}

We also see a relationship between the overall helium content and the binary fraction in the core of the cluster, as seen in Figure \ref{Y_Binaries}, using the observed binary fractions from \cite{Milone:2012}. As the helium abundances of the clusters decrease, the incidence of binaries increases. This trend continues until the binary fraction reaches about 40\%, after which the correlation appears to be weaker. However, there are few clusters beyond a 40\% binary fraction, and the errors tend to be larger.  

The binary fractions were determined by \citet{Milone:2012}, who noted that the more massive clusters (using absolute magnitude as a proxy for mass) had lower binary fractions. From Figure \ref{Y_Mabs}, we find that the less massive clusters have lower helium content, and Figure \ref{Y_Binaries} shows that these same clusters have higher binary fractions, consistent with \citet{Milone:2012}. While our observed trend between helium and binary fraction may be driven primarily by the anti-correlated mass-binary fraction, it may be important to consider what the binary fraction means with respect to helium, as it may inform the multiple population scenario. Because the binary fraction determinations and helium estimates derive from the same ACS photometry (\citealt{Sarajedini:2007}), they are expected to be correlated. Although our tests suggest that binaries are largely classified as field stars in our BASE-9 analysis and do not contribute significantly to the posterior distribution, it is possible that remaining binaries classified as cluster members could weight the analysis towards slightly lower helium abundances. Thus, the relationship between binary fraction and helium abundance requires further investigation.

Both Figures \ref{Y_Mabs} and \ref{Y_Binaries} indirectly suggest a relationship between the helium of a cluster and its mass. Thus, it appears that there is a preference for clusters of higher mass to also have a greater overall enhancement of helium. This relationship has been previously seen for a smaller sample of clusters (\citealt{Milone:2014}, \citealt{Milone:2015}), but we find that it holds true for the majority of Galactic globular clusters. A recent study by \cite{Lucatello:2015} showed that second generation stars, which are presumably more helium enhanced, have a lower incidence of binaries. We also find that the more enhanced clusters, which likely harbor a greater proportion of second generation stars, have fewer binary systems.

\subsection{Light Elements}\label{CNOabunds}

\begin{figure*}
\centering
\includegraphics[width=\textwidth]{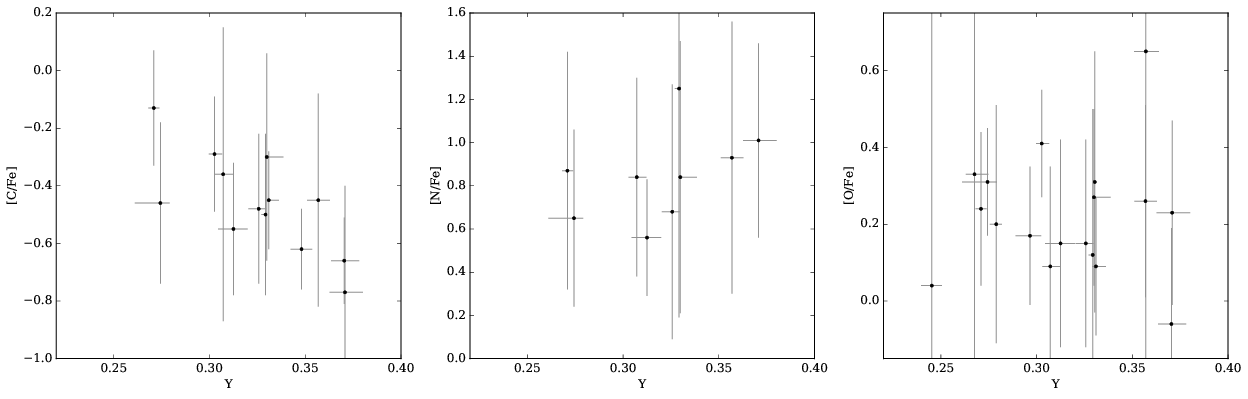}
\caption{From left to right, the helium fraction is plotted against the carbon to iron ratio, nitrogen to iron ratio, and oxygen to iron ratio. Carbon is negatively correlated with helium and nitrogen is positively correlated with helium. Oxygen appears uncorrelated with helium.}
\label{Y_CNO}
\end{figure*}

Additionally, we also find a possible correlation between helium and light element abundances. Understanding the connection between helium and the levels of carbon, oxygen, and nitrogen is a crucial step to disentangling the progenitors of the multiple population phenomena. Using the results from \citet{Roediger:2014} obtained from compiling spectral data from the literature, we compare our helium estimates to carbon, oxygen, and nitrogen abundances in Figure \ref{Y_CNO} for a small sample of clusters with spectroscopic measurements.

Despite the relatively small sample of clusters with CNO spectral observations and the large measurement errors, clusters with higher helium tend to have lower carbon abundances and greater nitrogen abundances. There is a strong correlation with a Pearson coefficient of --0.742 $\pm$ 0.004 for carbon and a moderate correlation (0.436 $\pm$ 0.240) for nitrogen. For oxygen, there is no apparent relationship with helium, with a correlation coefficient of essentially zero (0.060 $\pm$ 0.812).

For many proposed multiple population scenarios, the first generation of evolved stars provides processed material to the formation of the later generations of stars (\citealt{Renzini:2008}, \citealt{Gratton:2012}, \citealt{Bastian:2015}). The CNO cycle taking place in these progenitors would be expected to deplete carbon and oxygen abundances and boost nitrogen abundances, while the same processes produce helium. Our observations of these effects for carbon and nitrogen supports this scenario of multiple population formation, except we find little to no change in the oxygen abundances. While in Figure \ref{Y_CNO} we are comparing the average abundances for the cluster against the average helium value of the cluster, future work should aspire to compare the light element abundances and the helium enrichment of \emph{each} population. Although there are only a handful of clusters with measured abundances for carbon, oxygen, and nitrogen, we do not see a relationship between helium and total CNO.

\subsection{Red Giant Branch Bump}\label{RGBbump}

The red giant branch bump (RGBB) occurs due to the hydrogen-burning shell in the star moving outward during the first ascent of the RGB. Stars undergoing this event become temporarily brighter, until the hydrogen-burning shell reaches the chemical discontinuity near the convective envelope in the star. After this, the stars grow fainter again, causing a ``pile-up" in the CMD along the RGB. Awareness of the RGBB, clearly visible in many CMDs along the RGB, dates back to its prediction by \citet{Thomas:1967} and its confirmation by \citet{King:1985}. The RGBB of a cluster has long been predicted to be sensitive to helium, age, and metallicity (\citealt{Cassisi:1997}, \citealt{Bono:2001}, \citealt{Bjork:2006}, \citealt{Salaris:2006}, \citealt{DiCecco:2010}, \citealt{Cassisi:2011}, \citealt{Nataf:2013}). Such studies purport that an increase in initial helium abundance will increase the brightness of the RGBB.

\begin{figure*}
\includegraphics[width=\textwidth]{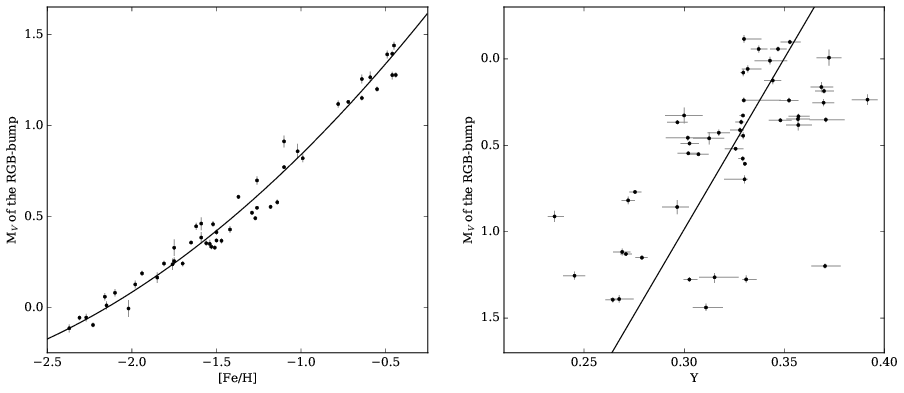}
\caption{The left panel shows the relationship of metallicity to the brightness of the RGBB. An error-weighted quadratic is fit to the data, shown as a solid line and detailed in Equation \ref{RGBBfit1}. The right panel demonstrates the apparent dependence of the RGBB magnitude on helium, though it is likely driven by a [Fe/H] dependence. We provide an error-weighted linear fit to the data regardless, shown as the solid line and detailed in Equation \ref{RGBBfit2}.}
\label{RGBB1}
\end{figure*}

Using measurements from \citet{Nataf:2013}, we compare the absolute V-magnitude of the RGBB to the metallicities and helium abundances in Figure \ref{RGBB1}. There is a clear trend in both, supporting predictions and previous work. The correlation between [Fe/H] and the RGBB magnitude is very strong, and appears to be non-linear. We fit an X and Y error-weighted quadratic to the data:

\begin{eqnarray}
M_{RGBB} =  0.161 (\pm0.030) [Fe/H]^2 +  \nonumber \\
1.240 (\pm0.088) [Fe/H] + 1.917(\pm0.064)
\label{RGBBfit1}
\end{eqnarray}

We also fit a linear model, again accounting for errors in both directions, to helium and absolute RGBB magnitude:

\begin{equation}
M_{RGBB} = -0.080 (\pm0.014) (Y \times 100) + 3.120 (\pm0.452)
\label{RGBBfit2}
\end{equation}

These fits are shown as solid lines in the panels of Figure \ref{RGBB1} ([Fe/H] on the left and helium on the right).

\begin{figure*}
\includegraphics[width=\textwidth]{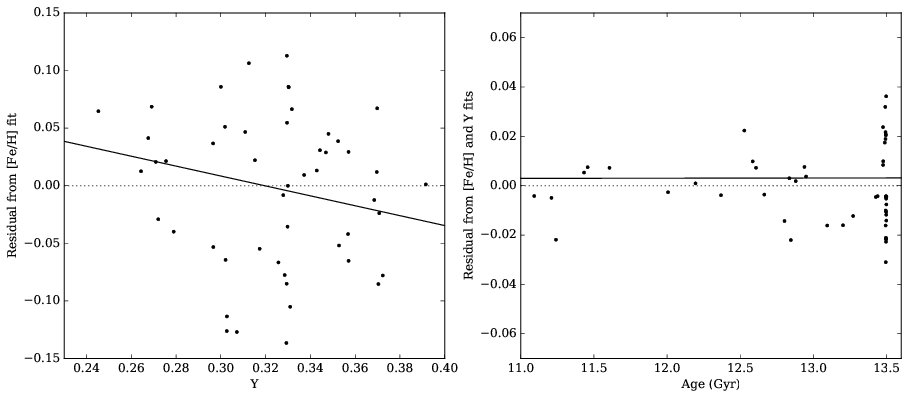}
\caption{The left panel shows the residual of the RGBB magnitude from the error-weighted fit to metallicity (observed - predicted) vs helium. The residuals have a median of 0.016 and a standard deviation of 0.072. A line is fit to the data with a slope of --0.429 $\pm$ 0.271 and an intercept of 0.137 $\pm$ 0.088. The right panel shows the residuals after the effect from helium is also removed from the data, consistent with zero. The residuals have a median of --0.004 and a standard deviation of 0.015.}
\label{RGBB2}
\end{figure*}

The correlation between helium and RGBB absolute magnitude is evident, with a weighted Pearson coefficient of --0.364$\pm$0.009. However, metallicity is clearly more important, and may drive a significant portion of the relationship between Y and RGBB magnitude. In the left panel of Figure \ref{RGBB2}, we plot the residuals from the fit between metallicity and RGBB magnitude against the helium of each cluster. The residuals have a median of --0.019 and a standard deviation of 0.068. An X and Y error-weighted linear model to the residuals results in a slope of --0.429 $\pm$ 0.271 and an intercept of 0.137 $\pm$ 0.088, suggesting that some effect due to helium remains. 
With the effect on the RGBB magnitude due to metallicity removed, our fit predicts that a change in 0.01 in Y results in a brighter RGBB by 0.008 magnitudes. This is similar to the result of \citet{Cassisi:1997}, who suggests that for the same increase in helium, the magnitude of $\Delta$(V$_{bump}$-V$_{TOP}$) becomes brighter by 0.011 magnitudes. \citet{Salaris:2006} also suggests a minor shift in the RGBB magnitude for an increase in helium.

Using our linear fit (shown in the left panel in Figure \ref{RGBB2}), we remove the remaining dependence of the RGBB magnitude on Y to examine any remaining residual after removing the dependencies on metallicity and helium. The remaining scatter is plot against the age of the clusters in the right panel of Figure \ref{RGBB2}. If we attempt to fit an error-weighted linear relation, the intercept is consistent consistent with zero. The remaining scatter could be from uncertainties in the distances or absorptions. The median residual is --0.004 with a standard deviation of 0.015; the standard deviation of the residuals is reduced by a factor of 4.5 after the removal of the remaining trend with helium.

Interestingly, we do not see any clear trends between the brightness of the RGBB and the light element abundances.

\subsection{RR Lyrae Periods}\label{RRL}

\begin{figure*}
\includegraphics[width=\textwidth]{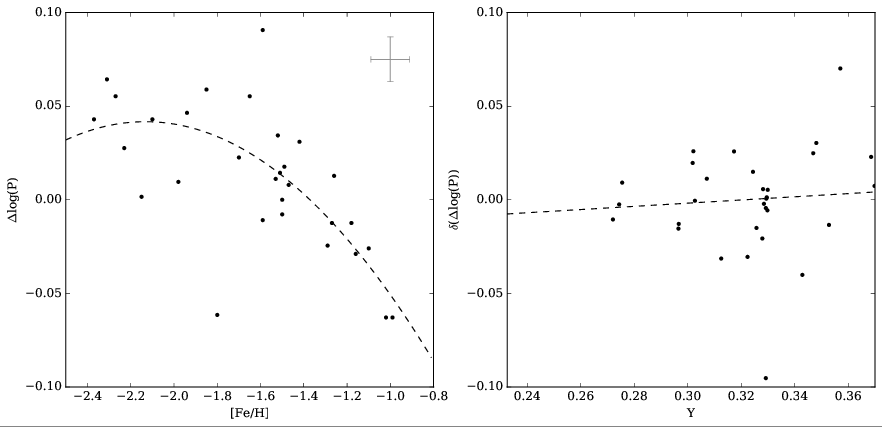}
\caption{Left: a recreation of the $\Delta$log(P) vs. metallicity diagram from \protect\cite{Lee:1994} with current data. We fit a quadratic to the relation (see Equation \ref{eq:LDZ1}). Right: the residuals from the quadratic fit plotted with helium, showing a modest, although not statistically significant, remaining trend.}
\label{fig:LDZ}
\end{figure*}

We also examine our results in the context of \cite{Lee:1994}, who explored the theoretical dependence of the horizontal branch morphology on parameters such as age, metallicity, and helium. \cite{Lee:1994} predicted that increased helium would lead to a greater $\Delta$log(P), which they defined as the difference between the fundamentalized average periods of clusters' RR Lyraes compared to M3. The fundamentalized period for each cluster was defined by (log(P$_f$) = log(P$_c$)+0.13), with P$_c$ being the average period of RR Lyrae c-type stars in each cluster (\citealt{Castellani:1987}). Consequently, the study concluded that the fundamentalized periods of clusters were \emph{not} affected by helium because the observational data did not follow the theoretical predictions, which suggested that an increase in helium would be reflected in an increased in $\Delta$log(P).

The fundamentalized periods of RR Lyrae type c observations from \cite{Castellani:1987} and the \cite{Harris:2010} metallicities are used to recreate Figure 10 from \cite{Lee:1994}, as seen in the left panel of Figure \ref{fig:LDZ}. In this panel, the fundamentalized RR Lyrae periods, with respect to the fundamentalized period of M3, are plot against the metallicities. A quadratic is fit to the relation

\begin{eqnarray}
\Delta Log(P) = -0.072(\pm0.024) [Fe/H]^2 \nonumber \\
- 0.308(\pm0.080) [Fe/H] - 0.286(\pm0.065)
\label{eq:LDZ1}
\end{eqnarray}

We exclude the two outliers, NGC 2419 (lower) and NGC 6333 (upper) in the fit. Leaving out these same two clusters, we find that residuals of Equation \ref{eq:LDZ1} are related to helium by the following relation:

\begin{equation}
Resid = 0.085(\pm0.143) Y - 0.027(\pm0.046)
\label{eq:LDZ2}
\end{equation}

Qualitatively, our results match the general trend predicted by the theoretical calculations from \cite{Lee:1994} that an enriched Y value leads to a greater $\Delta$log(P). However, the trend is not statistically significant, suggesting that this approach cannot provide any hard conclusions on the possible effects of helium on the horizontal branch morphology.

\subsection{Anomalous Clusters}

While most globular clusters harbor distinct populations that vary in helium and light element abundances, a smaller subset of clusters has been found to also show variations in [Fe/H] abundances and \emph{s}-process element abundances (\citealt{Bedin:2004, Gratton:2012, Marino:2015}). Although only a handful of clusters have been categorized as anomalous clusters, there are likely more that have not yet been the target of sufficient high-resolution spectroscopy to be identified as such.

Several anomalous clusters in the Milky Way overlap our cluster sample: $\omega$ Centauri (NGC 5139), NGC 1851, M22 (NGC 6656), M2 (NGC 7089), M54 (NGC 6715), and NGC 5286 (\citealt{Milone:2009, Milone:2013, Marino:2015}). Except for NGC 1851, the most metal-rich and the youngest of the anomalous clusters in our sample, the clusters have above average helium abundances. Otherwise, we do not notice anything particularly unusual about the characteristics of the anomalous clusters in our sample, probably due to only a rough sensitivity to the complex multiple population characteristics from visual photometry.


\section{Discussion}\label{Discussion}

We stress that our results are dependent on two primary assumptions. First, we assume the clusters analyzed have a single iron abundance and a single age. For some clusters with multiple populations this may not be the case. In a few clusters, ages of different populations of stars could vary and the internal abundance spread may be significant. However, for the handful of clusters where this may be the case, we expect that the internal variation in age is within the fitting errors.

Secondly, our methodology is dependent on the validity of the theoretical models, which are assumed to be accurate (\citealt{von-Hippel:2006, De-Gennaro:2009, Stein:2013, Hills:2015}). Our results, while precise, can therefore only be as accurate as the models. We note that the quoted uncertainties on the parameters derived herein represent the statistical uncertainty. This does not incorporate uncertainties in an astronomical sense that may arise from model misspecification, unincorporated effects, incorrect assumptions about physical processes, or observational systematics. Once models improve, the approach presented here will allow us to focus on absolute quantities, such as absolute helium abundances and absolute ages. In the meantime, the high precision results we obtain from BASE-9 fits with HST data allow us to perform numerous relative analyses.

The fitted helium values can be interpreted as ``average" helium values, which depend not only on the extent of the helium enhancement of stars in the cluster, but also the fraction of stars that are enhanced in helium. In this study, these two are degenerate. Further work is needed to disentangle these effects and study their individual influences on the correlations we see.

The relationships we find among the helium fraction and the light element abundances of carbon and nitrogen, and the binary fraction, are important avenues for further investigation with respect to the multiple population scenario. Ultraviolet observations of these clusters are now a common avenue for analyzing multiple populations. The ultraviolet wavelengths encode the relative abundances of the light elements carbon, nitrogen, and oxygen, rather than directly probing helium of a cluster (\citealt{Piotto:2015}). Thus, it is very important to understand any relationships between abundances of various light elements and the helium content. This is especially true as helium and $\alpha$-enhancement tend to have off-setting effects on the shape and location of theoretical isochrones. These effects are more dramatic in the ultraviolet than in the visual wavelengths, but in either case, are generally more exaggerated for the more metal-rich clusters.

The correlation between helium abundance and binary fraction is also of interest to multiple population research. In multiple population formation models for clusters that invoke multiple star formation episodes, it is often suggested that the first generation of stars was significantly ($/geq$ 10$\times$) more massive than the present-day cluster masses (\citealt{Bastian:2015, Cottrell:1981, Gratton:2004, DErcole:2008, Decressin:2007a}). The first generation of stars is then largely lost to the field star population. Previous work has suggested that first generation stars in clusters have higher binary fractions (up to 15\%) compared to later generations, which have only a few percent (\citealt{Milone:2008}, \citealt{Lucatello:2015}). Our results show the more massive clusters with a lower binary fraction tend to have more helium, as we may expect from stars belonging to the chemically enriched population. We also observe that the less enriched clusters appear to be less massive and have a higher incidence of binaries, as is predicted for first generation dominated clusters. Further investigation of the binary fraction with respect to helium may be able to shed more light on the different binary fractions in globular clusters (and their sub-populations) and field stars (\citealt{Milone:2008}, \citealt{DOrazi:2010}, \citealt{Gratton:2012}), thereby providing information on the source of the field star population.


\section{Conclusions}\label{Conclusion}

We analyzed 69 Galactic globular clusters to simultaneously determine their their precise relative ages, distances, extinctions, and helium fractions with statistical rigor. Based on the results, we reach the following conclusions:

1. We find a symmetric distribution of helium fractions for the clusters, ranging from approximately 0.235 to 0.409 with a peak at Y $\sim$ 0.321.

2. In the age-metallicity relation, there is a clear demarcation between two unique age-metallicity sequences, with older clusters being primarily located in the inner halo, and younger clusters in the outer halo. This suggests a multi-phase formation and evolution of the Milky Way, consistent with previous studies.

3. Correlations between the helium fractions of the clusters and both their absolute magnitudes and their binary fractions lend additional evidence to the suggestions that more massive clusters harbor higher helium contents.

4. Carbon and nitrogen abundances are found to be correlated with the overall helium fractions of several clusters. Although spectral measurements are still relatively sparse, it appears that higher helium is associated with a lower [C/Fe] ratio and a greater [N/Fe] ratio. We find no statistically significant relationship between helium and oxygen abundance.

5. The absolute magnitude of the RGBB and the helium fraction are related, as predicted from previous studies; a change of 0.01 in helium results in a brighter RGBB by 0.008 magnitudes.


\section*{Acknowledgments}
We are grateful to the referee, Raffaele Gratton, whose comments greatly improved the paper. Support for this work (proposal number GO-10775) was provided by NASA through a grant from the Space Telescope Science Institute which is operated by the Association of Universities for Research in Astronomy, Incorporated, under NASA contract NAS5-26555. This material is based upon work supported by the National Aeronautics and Space Administration under Grant NNX11AF34G issued through the Office of SpaceScience and through the University of Central Florida's NASA Florida Space Grant Consortium. DS work was supported by NSF grants DMS 1208791. DvD was partially supported by a Wolfson Research Merit Award (WM110023) provided by the British Royal Society and by Marie-Curie Career Integration (FP7-PEOPLE-2012-CIG-321865) and Marie-Skodowska-Curie RISE (H2020-MSCA-RISE-2015-691164) Grants both provided by the European Commission.


\newpage
\bibliographystyle{mn2e}
\bibliography{ACS_GCs}
\clearpage


\label{lastpage}

\end{document}